\documentclass{amsart}
\usepackage{graphicx,amssymb}
\theoremstyle{definition}

\theoremstyle{remark}

\newcommand{\eps}{\varepsilon}

\def\parb{\pmb{\partial}}
\def\be{\begin{equation}}
\def\ee{\end{equation}}
\def\beq{\begin{eqnarray}}
\def\eeq{\end{eqnarray}}
\def\lb{\label}
\def\d{\mathrm{d}}

\def\bigO{\mathcal{O}}

\def\a{\alpha}
\def\b{\beta}
\def\ab{{\alpha\beta}}
\def\z{\epsilon}    
\def\G{\Gamma}
\def\Gv{\Gamma v}
\def\ehat{\hat{\mathbf{e}}}
\def\Sp{\Sigma_+{}}

\def\la{\langle}
\def\ra{\rangle}

\def\udot{\dot{u}}

\def\tauf{\tau_{\rm fluid}}
\def\Hf{H_{\rm fluid}}
\def\geff{\gamma_{\rm eff}}
\def\gp{\gamma_\perp}
\begin{document}

\title{Fluid observers and tilting cosmology}%
\author[A.A. Coley, S. Hervik and  W.C. Lim]{A.A. Coley, S. Hervik, W.C. Lim}%
\address{Department of Mathematics \& Statistics, Dalhousie University,
Halifax, Nova Scotia,
Canada B3H 3J5}%
\email{herviks@mathstat.dal.ca, wclim@mathstat.dal.ca, \newline aac@mathstat.dal.ca}%

\date{\today}

\begin{abstract}

We study perfect fluid cosmological models with a constant
equation of state parameter $\gamma$ in which there are two
naturally defined time-like congruences, a geometrically defined
geodesic congruence and a non-geodesic fluid congruence. We
establish an appropriate set of boost formulae relating the
physical variables, and consequently the observed quantities, in
the two frames. We study expanding spatially homogeneous tilted
perfect fluid models, with an emphasis on future evolution with
extreme tilt. We show that for ultra-radiative equations of state
(i.e., $\gamma>4/3$), generically the tilt becomes extreme at late
times and the fluid observers will reach infinite expansion within
a finite proper time and experience a singularity similar to that
of the big rip. In addition, we show that for sub-radiative
equations of state (i.e., $\gamma < 4/3$), the tilt can become
extreme at late times and give rise to an effective quintessential
equation of state. To establish the connection with phantom
cosmology and quintessence, we calculate the effective equation of
state in the models under consideration and we determine the
future asymptotic behaviour of the tilting models in the fluid
frame variables using the boost formulae. We also discuss
spatially inhomogeneous models and tilting spatially homogeneous
models with a cosmological constant.

\end{abstract}

\maketitle

\section{Introduction}

In cosmology it is essential to specify a set of observers, or
rather a congruence of world-lines, from which observations are
made. Physical quantities, such as the Hubble parameter, depend on
the choice of congruence and consequently the Hubble parameter is
observer-dependent. In this paper, we will consider a universe
with two different congruences of observers, with respect to which
the interpretation of the Universe can be radically different.
Spatially homogeneous perfect fluid  universes in which the fluid
is not necessarily comoving with the spatially homogeneous
hypersurfaces are prime examples of this phenomenon, and have the
advantage that they are dynamical solutions to Einstein's field
equations and can model some features of our real Universe. For
example,  in these models the observers moving with the geometric
congruence can experience an ever-expanding non-inflating
universe, while observers moving with the matter congruence can
experience an inflating universe.

For spatially homogeneous (SH)  Bianchi cosmological models, the
universe is foliated into space-like hypersurfaces (defined by the
group orbits of the respective model) \cite{EM,DS1,DS2,BS}.  For
these perfect fluid models there are two naturally defined time-like vector
fields (i.e., congruences): the unit vector field, $n^{a}$, normal
to the group orbits and hence orthogonal to the surfaces of
transitivity (the 'geometric' congruence), and the four-velocity,
$u^{a}$, of the fluid (the 'matter' congruence); see
Figure 1. If $u^{a}$ is not aligned with $n^{a}$, the model is
called \emph{tilted} (and non-tilted or orthogonal otherwise)
\cite{KingEllis}. The geometric congruence is necessarily
geodesic, vorticity-free and acceleration-free. The matter
congruence , on the other hand, is not necessarily geodesic and
can have both vorticity and acceleration. 
{\footnote{Perfect fluids with non-zero pressure have non-zero 4-acceleration
when $u^{a} \neq n^{a}$.}}
Usually, the kinematical
quantities associated with the normal congruence $n^a$ of the
spatial symmetry surfaces, rather than the fluid flow $u^a$, are
used as variables (a comoving fluid description is presented in
Appendix A). Following \cite{KingEllis}, tilt variables $v^\alpha$
are introduced, so that in an orthonormal frame where
$n^a=(1,0,0,0)$, we have
\begin{equation}
  u^a=\frac{1}{\sqrt{1-v^2}}\left(1,v^1,v^2,v^3\right),
\end{equation}
where $v^2 \equiv v_{\alpha}v^{\alpha}, v^2 \leq 1$.

\begin{figure}
\centering
\includegraphics[width=8cm]{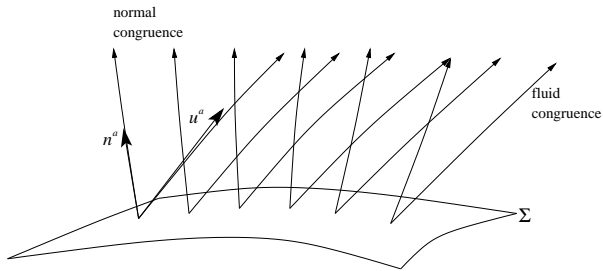}
\caption{The normal and the fluid congruences defined by the
vector fluids $n^a$ and $u^a$, respectively. The space-like
hypersurfaces $\Sigma$ are orthogonal to $n^a$ and are defined for
each Bianchi types by the group orbits of the respective Lie
group. }
\end{figure}
 We will  assume a perfect-fluid matter source with
$p=(\gamma-1)\mu$ as equation of state, where $\mu$ is the energy
density, $p$ is the pressure, and $\gamma$ is a constant.
Causality then requires $\gamma$ to be in the interval
$0\leq\gamma\leq2$. A positive cosmological constant may also be
included in the models. Such SH tilted cosmologies with a
$\gamma$-law perfect fluid source have been studied by a number of
authors. The tilted Bianchi type II model was studied in
\cite{HBWII}, the type V model in
\cite{Shikin,Collins,CollinsEllis,HWV,Harnett}, type VI$_0$ in
\cite{hervik,coleyhervik,BHtilted}, types IV and VII$_h$ in
\cite{CH,HHC}, a subset of the type VI$_h$ models in \cite{CH},
irrotational type VII$_0$ models in \cite{CH,LimDW}, general
tilted type VII$_0$ models in \cite{HHLC}, and tilted type VIII
models in \cite{HLim}.

It is known that the tilt can become extreme asymptotically (i.e.,
$v^2 \rightarrow 1$). Although it is possible for tilt to be
extreme both to the past and to the future, we shall focus on the
late-time behaviour of the tilted Bianchi models here. The extreme
tilt future attractors in SH models (without a cosmological
constant) are summarized in Table 1 (using the notation of
\cite{hervik} -- \cite{HLim}). In particular, let $T$ be the
proper time as measured along the fluid congruence. The fluid
congruence is usually inconvenient for computation, and so a
second congruence utilizing a temporal gauge called the separable
volume gauge \cite{UEWE} is used (e.g., the orthogonal frame for
SH models). Let $t$ and $H$ be the proper time and the Hubble
scalar associated with this congruence. The separable volume gauge
allows a convenient computational time variable $\tau$, called the
dynamical time, to be defined globally for the entire congruence
of world-lines: \be \lb{def_tau}
    \frac{\d\tau}{\d t} = H\, .
\ee

\begin{table}
\centering
\begin{tabular}{|c|c|c|l|}
\hline
 Bianchi & & Extreme    &  Other sinks\\
 model & Matter & tilt sink  & in set \\ \hline \hline
  II & $\frac{14}{9}<\gamma<2$ & $\mathcal{E}$(II) & No\\
        \hline
IV & $\frac{6}{5+2\Sigma_+}<\gamma<2$, $-\frac 14<\Sigma_+<0$ & $\widetilde{\mathcal{L}}_-$(IV) & Yes for $\gamma<\frac{4+\Sigma_+}{3}$ \\
         & $\frac{3}{2-\Sigma_+}<\gamma<2$, $-\frac 12<\Sigma_+<-\frac 14$ & $\widetilde{\mathcal{E}}_-$(IV) & No \\
        \hline
V & $\frac 65<\gamma<2$ & ${\mathcal{M}}_-$ & Yes for $\frac 65<\gamma<\frac 43$ \\
        \hline
VI$_0$ & $\frac 65<\gamma<2$ & ${\mathcal{E}}$(VI$_0$) & No \\
        \hline
VI$_h$ & ${\tilde{\gamma}_h}<\gamma<2$ [*] & [*] & [*] \\
\hline
VII$_0$ & $\frac 43<\gamma<2$ & ${\widetilde{P}_4}$(VII$_0$) & No \\
        \hline
VII$_h$ &$\frac{6}{5+2\Sigma_+}<\gamma<2$, $-\frac 14<\Sigma_+<0$ & $\widetilde{\mathcal{L}}_-$(VII$_h$) & Yes ($\mathcal{P}$) for $\gamma<\frac{4+\Sigma_+}{3}$ \\
         & $\frac{3}{2-\Sigma_+}<\gamma<2$, $-\frac 12<\Sigma_+<-\frac 14$ & $\widetilde{\mathcal{E}}$(VII$_h$) & No \\
\hline
VIII &    $1<\gamma<2$ & $\widetilde{E}_1$(VIII) & No \\
\hline
\end{tabular}
\caption{Summary of extremely tilted future attractors. [*] There
are extremely tilted attractors in the Bianchi VI$_h$ models, but
not all of the sinks have been determined in this case to date;
for $-1<h<0$, $h\neq -1/9$, we have that ${\tilde{\gamma}_h}< 6/5$
(for $h<-1$, we have that ${\tilde{\gamma}_h}< 4/3$).}

\end{table}

For an individual fluid world-line, along which the spatial
coordinates are held fixed, we have the following relation between
$T$ and $t$: \be \lb{dT_dt}
    \d T = \sqrt{1-v^2} \, \d t\, .
\ee We can now express the proper time elapsed, $\Delta T$, along
this fluid world-line as an integral over $\tau$: \be \lb{def_T}
    \Delta T = T_2-T_1=\int_{t_1}^{t_2}\sqrt{1-v^2}\, \d t
    = \int_{\tau_1}^{\tau_2}\frac 1H\sqrt{1-v^2} \, \d \tau\, .
\ee Physically, $\tau$ is directly related to the length scale
$\ell$; $H = \frac{1}{\ell}\frac{\d \ell}{\d t}$ and
(\ref{def_tau}) imply that $\ell = \ell_0(x^i) e^\tau$. As $\ell
\rightarrow \infty$, so does $\tau$. In this paper we are
interested in whether $\Delta T$ is finite or infinite as $\tau_2
\rightarrow \infty$. If $\Delta T$ is finite and $H \rightarrow
\infty$ as $\tau_2\rightarrow \infty$, then the congruence is
future incomplete: the fluid observers will reach infinite
expansion within finite proper time. Therefore, in spite of the
fact that these spacetimes are future geodesically complete
\cite{Rendall}, the world-lines defined by the fluid congruences
can approach null infinity so quickly that the proper time as
measured by the fluid observer is \emph{finite}.

In this paper, we study perfect fluid cosmological models in which
there are two naturally defined congruences, a geometrically
defined geodesic congruence and a non-geodesic fluid congruence.
In Appendix B we establish an appropriate set of boost formulae
relating the physical variables in the two congruences. We study
expanding spatially homogeneous tilted perfect fluid models, with
an emphasis on future evolution with extreme tilt. In order to set
the stage, we shall first discuss the Bianchi type V model in some
detail. We shall then discuss spatially homogeneous models with a
cosmological constant and various models (Bianchi types VII$_0$,
VII$_h$ and VIII) in the absence of a cosmological constant. We
then briefly discuss general spatially inhomogeneous models with a
cosmological constant. We find that for $\gamma
> 4/3$, expanding spatially homogeneous tilting
universes come to a violent end after a finite proper time,
reminiscent to that of cosmological models with a 'phantom energy'
equation of state parameter which end in a {\it big rip}. We also
find that for $\gamma < 4/3$) the tilt can become extreme at late
times and give rise to an effective quintessential equation of
state.

To establish the potential connection with phantom cosmology and
quintessence, we determine the effective equation of state in the
models under consideration as seen by the fluid observer. 
In order to investigate the future
asymptotic behaviour of the cosmological models we then determine
the asymptotic behaviour of the tilting models in the fluid frame
variables using the boost formulae derived in Appendix B. We also
study the asymptotic dynamical behaviour of a class of LRS Bianchi
type V solutions into the past, and show that they can be extended
along the fluid congruence. We conclude with a discussion, with an
emphasis on the physical interpretation of the models.

\section{Bianchi type V}

The irrotational Bianchi type V perfect fluid models were studied
in \cite{HWV}; it was shown that the tilt can become extreme (in a
finite time as measured along the fluid congruence). The global
structure of Bianchi V  models was studied in more detail in
\cite{CollinsEllis}. It was shown that for models with $4/3 <
\gamma \leq 2$, to the future, as $T$ approaches some finite
limiting value, the fluid world-lines become null (with respect to
the geometric congruence), and the expansion, the shear, the
acceleration, and the length scale $\ell$ all diverge but the
matter density (and curvature scalars) tend to zero; i.e., the
models end at a 'conformal singularity' (this peculiar behaviour
is depicted in Figs 7(ii) and 8 in \cite{CollinsEllis};
dynamically an observer is accelerating so quickly that their
world-line becomes null near null infinity in finite proper time,
signifying the onset of structural instability in the field
equations in these models).

This peculiarity occurs for ultra-radiative equations of state.
Such an equation of state may not be physically appropriate at
late times, although in \cite{BarTip} it was argued that the
future is dynamically radiation dominated (but with an always
sub-radiative equation of state). This perhaps indicates that
models (or rather the governing equations) with an ultra-radiative
equation of state may not be well-posed at late times (i.e.,
structurally unstable). However, we shall show later that a
similar type of physical behaviour may occur for particular SH
universes with a sub-radiative equation of state.

\subsection{Analysis}

We shall first show that the fluid proper time is finite as the
solution approaches the equilibrium point $M^-$  for the case $4/3
< \gamma < 2$ in the absence of a cosmological constant (see Fig.
10 in \cite{HWV}). We shall make use of the evolution equations
(denoted by HW) in \cite{HWV}. There, the equilibrium point $M^-$
is described by
\be
    \Sigma_+ = \Sigma_- =0\, ,\quad
    A=1\, ,\quad
    v=-1\, ,
\ee with the density parameter $\Omega =0$ from HW(2.13) and the
deceleration parameter $q = 0$ from HW(2.14). By linearizing 
equations HW(2.10)
and HW(2.11) about $M^-$, we heuristically obtain
the decay rates as $\tau \rightarrow \infty$ for
the Hubble scalar $H=\theta/3$ and the quantity $\sqrt{1-v^2}$:
\be
    H \approx H_0 e^{-\tau}\, ,\quad
    \sqrt{1-v^2} \approx (\sqrt{1-v^2})_0 \exp\left(
    -\frac{5\gamma-6}{2-\gamma} \tau \right)\, ,
\ee which have been confirmed by our numerical simulations.
Finally, it follows from (\ref{def_T}) that the fluid proper time
in the future asymptotic regime is
\be
    \Delta T \approx \int_{\tau_0}^\infty \frac{(\sqrt{1-v^2})_0}{H_0}
    \exp\left( -\frac{2(3\gamma-4)}{2-\gamma} \tau \right) \ \d
    \tau\,,
\ee which is finite for $4/3 < \gamma < 2$, as claimed. This is
consistent with the analysis of Collins and Ellis \cite[second
table in Fig. 7]{CollinsEllis}.

\subsection{Cosmological constant}

Spatially homogeneous cosmological models with a positive
cosmological constant were investigated using dynamical systems
methods in \cite{GE}, extending the tilted LRS Bianchi type V
analysis of \cite{HWV} to the $\Lambda\neq0$ case. Unlike the case
of orthogonal Bianchi models (including type IX), in which the de
Sitter point is the only new equilibrium point, when extending
these models from the $\Lambda=0$ case tilted models have
additional equilibrium points, namely the de Sitter point with
extreme tilt for $\gamma \neq 4/3$, and with arbitrary tilt for
$\gamma = 4/3$.

In the LRS case, the field equations and conservation equations
give rise to a low dimensional constrained dynamical system when
written in terms of dimensionless variables. Focusing on the
future evolution of these models, all of the sinks were
determined. The conventional de Sitter solution is a sink for
$\gamma<4/3$. The only sinks in the case $\gamma>4/3$ are the de
Sitter points with extreme tilt ($v=\pm1$). The threshold value
$\gamma=4/3$ is particularly interesting, as a non-zero
cosmological constant then implies $v^\prime=0$ for {\em any}
value of $v$. In other words, for each value of the tilt variable
$v$, there is a de Sitter equilibrium point \cite{GE}.

Therefore, in general, for Bianchi type V models with
$\gamma>4/3$, the tilt becomes extreme at late times. The fluid
motion is no longer orthogonal to the surfaces of homogeneity, and
the rest spaces of an observer comoving with the fluid need not be
spatially homogeneous and the model will not seem to isotropize.
(When following the normal congruence, on the other hand, the
fluid will no longer look perfect.)

It has been shown that expanding non-type-IX Bianchi models with a
positive cosmological constant isotropize to the future (cosmic no
hair theorem) \cite{Wald}. This result applies to tilted models,
but the isotropization is with respect to the congruence normal to
the homogeneous symmetry surfaces -- not the fluid congruence
\cite{RaychaudhuriModak}. Thus in the  Bianchi type V models the
the space-time generically becomes de Sitter-like, in accordance
with the cosmic no hair theorem, but since the tilt does not die
away, isotropization of the cosmology does not occur with respect
to the fluid congruence. It may be possible to extend beyond the
apparent limiting point to a region with time-like symmetries, as
discussed in \cite{CollinsEllis}.

\section{Spatially Homogeneous Models}

In general, SH models are not asymptotically isotropic.

\subsection{Isotropic Asymptotia}

The spatially homogeneous models isotropize to the future in the
presence of a positive cosmological constant \cite{Wald}.

\subsubsection{Cosmological constant}
Let us investigate a SH model with a cosmological constant and a
tilted perfect fluid with $p=(\gamma-1)\mu$. Consider the
asymptotic behaviour of the following solution which corresponds to
an equilibrium point of the system of governing equations:\\
Solution: de Sitter (dS) with an extremely tilted perfect fluid. \\
Asymptotic behaviour \cite{leuw}: $H=H_0$, $\sqrt{1-v^2}\approx
(\sqrt{1-v^2})_0
\exp\left(-\frac{3\gamma-4}{2-\gamma}\tau\right)$. \\
This gives for $\Delta T$: \be \Delta T\approx
\int_{\tau_0}^\infty \frac{(\sqrt{1-v^2})_0}{H_0}
    \exp\left(- \frac{3\gamma-4}{2-\gamma} \tau \right) \ \d \tau\, ,
\ee which is finite for $4/3 < \gamma < 2$. This solution is the
future attractor for $4/3<\gamma<2$ and therefore represents the
generic behaviour for this model.

\subsubsection{Intermediate behaviour}
In the absence of a cosmological constant, in general SH models
are not asymptotically isotropic. However, these models can spend
a long time close to a Friedmann model corresponding to a saddle
point \cite{DS1}. Therefore, let us consider one non-tilting
perfect fluid with $p_{\perp}=(\gp-1)\mu_{\perp}$ and one
tilting fluid with $p=(\gamma-1)\mu$. We assume that the
non-tilting fluid is dominant and has $\gp<\gamma$.
Consider the equilibrium point:\\
Solution: Friedmann $F$, with $H\approx H_0 \exp(-\frac32\gp\tau)$
with
extremely tilted second fluid.\\
Asymptotic behaviour: $\sqrt{1-v^2}\approx (\sqrt{1-v^2})_0 \exp\left(-\frac{3\gamma-4}{2-\gamma}\tau\right)$. \\
This gives for $\Delta T$: \be \Delta T\approx
\int_{\tau_0}^\infty \frac{(\sqrt{1-v^2})_0}{H_0}
    \exp\left(- \frac{3\gamma-4-\frac 32\gp(2-\gamma)}{2-\gamma} \tau
\right) \ \d \tau.
\ee In the class of SH models $F$ is a saddle for $2/3<\gp<2$
and represents possible intermediate behaviour. This means that
for solutions spending a finite (but arbitrarily long) time close
to the saddle $F$, exhibiting quasi-isotropic behaviour consistent
with observations  \cite{DS1}, the Hubble parameter as measured by
the fluid, $H_{\text{fluid}}$, can become arbitrarily large. Of
particular interest are the cases when the non-tilting fluid is
dust ($\gp=1$) or radiation ($\gp=4/3$) for which $\Delta T$
is finite for $\gamma>14/9$ and $\gamma>8/5$, respectively.

\subsection{Anisotropic Asymptotia}

There will be different asymptotic end-states depending on Bianchi
types (where the criteria may depend on $\gamma$). In all of the
cases below we have one tilted perfect fluid.

\subsubsection{Bianchi type VII$_0$}

The investigation of the type VII$_0$ model is important  for
several reasons.  First, the type VII$_0$ model is the most
general spatially homogeneous model allowing for the flat
Friedmann-Robertson-Walker (FRW) model as a special case.  Both
the Bianchi type I model and the VII$_0$ generalise the flat FRW
model; however, since the Bianchi type I perfect fluid model does
not allow for tilt, only the Bianchi type VII$_0$ model can be
used to investigate the effect of tilt on the evolution of the
universe close to flatness. In addition, the type VII$_0$ model
plays an important role in the Bianchi hierarchy. For example, the
type VII$_0$ is a special limit of both the 'most general' type
VIII and type IX models. Also, unlike the other Bianchi models of
solvable type, the type VII$_0$ state space is unbounded. The type
VII$_0$ model is not asymptotically self-similar to the future
because the unbounded growth of one of the curvature variables
leads to oscillations in the shear and the curvature variables
\cite{DS1}.

The Bianchi type VII$_0$ model with a tilted $\gamma$-law perfect
fluid was studied in \cite{HHLC} using expansion-normalised
variables and utilizing a formalism adapted to the time-like
geodesics orthogonal to the hypersurfaces defined by the type
VII$_0$ group action. In particular, the future asymptotic states
were determined. Regarding the tilt velocity, it was shown that
for fluids with $\gamma<4/3$ (which includes the important case of
dust, $\gamma=1$) the tilt velocity tends to zero at late times,
while for a radiation fluid, $\gamma=4/3$, the fluid is tilted and
its vorticity is dynamically significant at late times. For fluids
stiffer than radiation ($\gamma>4/3$), the future asymptotic state
is an extremely tilted spacetime with vorticity.

For $4/3<\gamma<2$, the solution corresponding to the sink
\cite{HHLC} has the asymptotic form \be H\approx H_0e^{-2\tau},
\quad
\sqrt{1-v^2}\approx(\sqrt{1-v^2})_0\exp\left(-\frac{3\gamma-4}{2-\gamma}\tau\right).
\ee This implies,
\begin{equation}
\Delta T\equiv \int_{\tau_0}^{\infty}\frac 1H\sqrt{1-v^2}{\d
\tau}\approx
\int_{\tau_0}^{\infty}\frac{(\sqrt{1-v^2})_0}{H_0}
\exp\left(-{\frac{5\gamma-8}{2-\gamma}\tau}\right)\d \tau,
\end{equation}
which corresponds to the proper time of an observer from
$\tau=\tau_0$ to $\tau=\infty$ following the fluid congruence,
there is a change of behaviour at $\gamma=8/5$. For models with
$\gamma \leq 8/5$, we find that this integral diverges. However,
for fluids with $\gamma>8/5$, we find that this integral is
finite. Since asymptotically at late times a fluid with
$\gamma>8/5$ will be extremely tilted, this means that the fluid
will reach future null infinity in finite proper time. This is
consistent with the work of \cite{BarTip} \footnote {In
particular, the dynamical future of radiation dominated Bianchi
VII$_0$ models, regarded as homogeneous perturbations of flat
Friedmann model, was studied in \cite{BarTip}. At late times it
was found that $v \sim \ell ^{( 3\gamma-4)}$, ${\omega/H}^2 \sim
\ell ^{( 9\gamma-10)}$, signifying a growing vorticity
instability. It was consequently found that the flat radiation
filled Friedmann model is unstable to the development of
vorticity.}. As discussed earlier, a similar phenomenon occurs for
the LRS Bianchi type V model with $\gamma>4/3$
\cite{CollinsEllis}.

\subsubsection{Bianchi type VII$_h$}
For this model the appropriate solutions are vacuum plane waves with extreme tilt (see \cite{CH,HHC} for details). There are two possible asymptotic states depending on a parameter $\Sigma_+$:
\begin{enumerate}
\item{} $\widetilde{\mathcal{L}}_-(VII_h)$: Stable for $6/(5+2\Sigma_+)<\gamma<2$, $-1/4<\Sigma_+<0$. \\
Asymptotic behaviour: $H\approx H_0e^{-(1-2\Sigma_+)\tau}$,
\[ \sqrt{1-v^2}\approx (\sqrt{1-v^2})_0 \exp\left(-\frac{2(5\gamma-6+2\gamma\Sigma_+)}{2-\gamma}\tau\right).\]
This gives for $\Delta T$:
\be
\Delta T\approx \int_{\tau_0}^\infty \frac{(\sqrt{1-v^2})_0}{H_0}
    \exp\left(- \frac{2(3\gamma-4+2\Sigma_+)}{2-\gamma} \tau \right) \ \d \tau.
\nonumber\ee
\item{} $\widetilde{\mathcal{E}}(VII_h)$: Stable for $3/(2-\Sigma_+)<\gamma<2$, $-1/2<\Sigma_+<-1/4$. \\
Asymptotic behaviour: $H\approx H_0e^{-(1-2\Sigma_+)\tau}$,
\[ \sqrt{1-v^2}\approx (\sqrt{1-v^2})_0 \exp\left(-\frac{2(1-2\Sigma_+)(2\gamma-3-\gamma\Sigma_+)}{3|\Sigma_+|(2-\gamma)}\tau\right).\]
This gives for $\Delta T$:
\be
\Delta T\approx \int_{\tau_0}^\infty \frac{(\sqrt{1-v^2})_0}{H_0}
    \exp\left(- \frac{(1-2\Sigma_+)^2(2\gamma-3)}{3|\Sigma_+|(2-\gamma)} \tau \right) \ \d \tau.
\nonumber \ee
\end{enumerate}
In the case $-1/2<\Sigma_+<-1/4$, the asymptotic state is actually
a closed orbit, so in calculating the decay rates we have
considered the appropriate averaged value as described in
Section 4 of \cite{HHC}. From the above expressions $\Delta T$ is
finite whenever $\gamma>(4-2\Sigma_+)/3$ for $-1/4<\Sigma_+<0$ and
$\gamma>3/2$ for $-1/2<\Sigma_+<-1/4$. However, since for a given type VII$_h$ model any $-1/2<\Sigma_+<0$
is possible, we have that: \\
\emph{For tilting perfect fluid models of Bianchi type VII$_h$
with equation of state parameter $4/3<\gamma<2$, there exists a
set of solutions of non-zero measure for which $\Delta T$ is
finite.}

\subsubsection{Bianchi type VIII}
These models are asymptotically extremely tilted for $1<\gamma<2$ and the
asymptotic solution is an extremely Weyl-curvature dominated model (in
the language of \cite{BHWeyl}). The solutions tend to a vacuum state but
do so rather slowly: $\mu/H^2\propto
1/\tau$. We refer to \cite{HLim} for details. \\
Asymptotic behaviour ($1<\gamma<2$): $H\approx H_0\tau^{\frac 14}\exp\left(-\frac 32\tau\right)$,
\[ \sqrt{1-v^2}\approx (\sqrt{1-v^2})_0 \tau^{\frac{1}{2(2-\gamma)}}\exp\left(-\frac{3(\gamma-1)}{2-\gamma}\tau\right).\]
This gives for $\Delta T$:
\be \Delta T\approx \int_{\tau_0}^\infty \frac{(\sqrt{1-v^2})_0}{H_0}
    \tau^{\frac{\gamma}{4(2-\gamma)}}
    \exp\left(- \frac{3(3\gamma-4)}{2(2-\gamma)}\tau\right) \d\tau,
\ee
which is finite for $4/3<\gamma<2$.

\section{General Spatially Inhomogeneous Models with a cosmological constant}

It is known that for general spatially inhomogeneous perfect fluid
models with a cosmological constant, the de Sitter solution with
extreme tilt%
\footnote{Here the tilt refers to the fluid tilt with respect to a
congruence whose acceleration tends to zero.} is locally stable
for $4/3 < \gamma < 2$ \cite{leuw,Rendall}. The de Sitter
asymptotic expansion is given in \cite{leuw}. From Equations
(3.43) and (3.28) of \cite{leuw}, we have that \be
        H \approx H_0 \, ,\quad
        \sqrt{1-v^2} \approx (\sqrt{1-v^2})_0 \exp \left(
        -\frac{3\gamma-4}{2-\gamma}\tau\right)\, ,
\ee
and thus
\be
        \Delta T \approx \int_{\tau_0}^\infty \frac{(\sqrt{1-v^2})_0}{H_0}
        \exp\left( -\frac{3\gamma-4}{2-\gamma} \tau \right) \ \d \tau\,
, \ee which is finite for $4/3 < \gamma < 2$, as required.

A comment on the physical ramification of this result is in order.
We note that for $4/3<\gamma<2$, the generic behaviour is
$v\rightarrow 1$, which implies that inflation does not isotropise
an ultra-radiative fluid. Even if inflation is turned off after a
certain number of $e$-foldings, $H_{\text{fluid}}$ can become
arbitrary large. However, as noted above, this result does not
contradict the cosmic no-hair theorem \cite{Wald}, since
generically the space-time is dynamically de Sitter-like (in the
normal frame).

\section{Effective equation of state}

We observe that the future dynamical evolution of the models under
investigation (with $\gamma > 4/3$) is reminiscent to that of
cosmological models with a constant equation of state parameter
$\gamma_{\mathrm{eff}} < 0 $, dubbed 'phantom energy', in which
expanding universes end in a singularity called the {\it big rip}
after a finite proper time \cite{phantom}. To establish the
connection with physics let us determine the effective equation
of state, as measured by the fluid observer, in the models under consideration.

For convenience, we first present the generic boost formula (see
Appendix B) for the Hubble scalar: \be
    \Hf = \Gamma B H \, ,
\ee where \be
        B = 1   - \tfrac23 v^\alpha A_\alpha
        + \tfrac13  v^\alpha \dot{U}_\alpha
                + \tfrac13 \G^{-1} \parb_0 (\G)
                + \tfrac13  \G^{-1} \parb_\alpha (\Gv^\alpha)\ .
\ee A fluid version of (\ref{def_tau}) can be defined along an
individual fluid world-line: \be
    \frac{\d \tauf}{\d T} = \Hf\, .
\ee Then the two dynamical times are related by \be
    \d \tauf = B\, \d \tau\, .
\ee We note that a finite  $\Delta T$ implies the divergence of
$H_{\text{fluid}}$. Indeed, in the examples above $H_{\rm fluid}$
diverges.

Let us now consider that the fluid observer interprets the
Universe as a flat Friedmann universe with an effective equation
of state. We define the effective density $\mu_{\rm eff}$ and
effective pressure $p_{\rm eff}$  for the fluid frame by writing
the Friedmann equation as $0 = 3 H_{\rm fluid}^2 - \mu_{\rm
eff}$, and the Raychaudhuri equation as $\mathbf{e}_0 H_{\rm
fluid} = - H_{\rm fluid}^2 - \frac16(\mu_{\rm eff} + 3p_{\rm
eff})$, as is standard. Equivalently, expressed in terms of
Hubble-normalized variables, we have $\Omega_{\rm eff} = 1$, $
q_{\rm fluid} = \tfrac12 (\Omega+3P)_{\rm eff}$. We define the
effective equation of state parameter $\geff$ as \be
    \geff = \frac{P_{\rm eff}}{\Omega_{\rm eff}} + 1\, ,
\ee 
so that $\geff$ is given by \be
    \geff = \tfrac23(q_{\rm fluid} + 1)\, .
\ee If the limit of $q_{\rm fluid}$ as $\tau \rightarrow \infty$
exists, then it can be read of from the leading exponent of
$H_{\rm fluid}$ since \be
    H_{\rm fluid} \approx (H_{\rm fluid})_0
                e^{-(q_{\rm fluid}+1)\tauf}\, ,
\ee
and $\tauf = B \tau$ in the limit.
Thus if $H_{\rm fluid} \approx (H_{\rm fluid})_0 e^{k\tau}$, then
$q_{\rm fluid} + 1 = -k/B$ and \
\be
    \geff = -\frac{2k}{3B}\, .
\ee

For the LRS Bianchi type V models \footnote{The effective equation
of state in the LRS Bianchi type V models can be calculated
directly in fluid-frame variables (see Appendix \ref{AppV}). }, on
the approach to $M^-$, we have that\be
    k = \frac{2(3\gamma-4)}{2-\gamma}\, ,\quad
    B = \frac{4}{3(2-\gamma)}\, ,
\ee
and so
\be
    \geff = -(3\gamma-4)
        < 0 \quad \text{for $\gamma > \tfrac43$;}
\ee
i.e., on approach to $M^-$ the ultra-radiative fluid
effectively behaves like a phantom energy in an isotropic and
spatially flat spacetime%
\footnote{Note that for $\gamma = \frac{4}{3}$, we obtain an
effective equation of state corresponding to a cosmological
constant. However, in this case the future attractors are not of
extreme tilt and the asymptotic decay rates will differ.}.

Similarly, for the de Sitter asymptotic state we have that \be
    k = \frac{3\gamma-4}{2-\gamma}\, ,\quad
        B = \frac{2}{3(2-\gamma)}\, ,
\ee
and coincidentally
\be
        \geff = -(3\gamma-4)
        < 0 \quad \text{for $\gamma > \tfrac43$.}
\ee This is the general case in inhomogeneous cosmological models
with a cosmological constant. As the de Sitter asymptotic state is
approached the ultra-radiative perfect fluid effectively behaves
like a phantom energy in an isotropic and spatially flat
spacetime. Note that inflation does not stop the big rip from
occurring.

Let us now consider the SH models discussed above. For the Bianchi
type VIII example, we have that
\be
        k = \frac{3(3\gamma-4)}{2(2-\gamma)}\, ,\quad
        B = \frac{1}{2-\gamma}\, ,
\ee and again \be \lb{28}
        \geff = -(3\gamma-4)
        < 0 \quad \text{for $\gamma > \tfrac43$.}
\ee We also note that for the Bianchi VIII model the future
attractor is of extreme tilt for $\gamma > 1 $, and so $0< \geff <
1$ for $1 < \gamma < \tfrac43 $.

For the Bianchi type VII$_0$ example, we have \be
        k = \frac{5\gamma-8}{2-\gamma}\, ,\quad
        B = \frac{2}{3(2-\gamma)}\, ,
\ee
and
\be
        \geff = -(5\gamma-8)
        < 0 \quad \text{for $\gamma > \tfrac85$.}
\ee We attribute this difference from the previous cases to the
fact that Bianchi type VII$_0$ models are geometrically more
special than Bianchi type VIII models. For the first Bianchi type
VII$_h$ example, we have that\be
        k = \frac{2(3\gamma-4+2\Sigma_+)}{2-\gamma}\, ,\quad
        B = \frac{4(1+\Sigma_+)}{3(2-\gamma)}\, ,
\ee
and
\be
        \geff = - \frac{3\gamma-4+2\Sigma_+}{1+\Sigma_+}
    < 0 \quad \text{for $\gamma > \tfrac43 -\tfrac23\Sigma_+$,}
\ee where $\Sigma_+$ satisfies $ -\tfrac14 < \Sigma_+ < 0$. And
for the second Bianchi type VII$_h$ example, we have that \be
        k = \frac{(1-2\Sigma_+)^2(2\gamma-3)}{3|\Sigma_+|(2-\gamma)}\, ,
\ee and $B$ is a complicated constant. But since $B$ is always
positive, we have that $\geff < 0$ for $\gamma > \tfrac32$.
Similarly, for the two-fluid example, we have $\geff < 0$ for
$\gamma > \tfrac{2(4+3\gp)}{3(2+\gp)}$. In the generic Bianchi
type VIII example the ultra-radiative perfect fluid effectively
behaves like a phantom energy. In the more special examples, the
requirement is that the equation of state must be equal to or
higher than that of radiation.

In order to compare the future asymptotic behaviour with the
behaviour of various types of big rip singularities
\cite{phantom,BarTsa}, in the next section we shall determine the
asymptotic behaviour of the tilting models in the fluid frame
variables.

\subsection{Quintessence}

We also see from above that there are models which have extreme
tilt asymptotically, but for $\gamma \leq \frac{4}{3}$ the
effective equation of state is that of a cosmological constant
($\geff=0$) or that of quintessence \cite{quint}
($0<\geff<\frac{2}{3}$). If $\gamma \leq \frac{4}{3}$, then the
future attractor may not be of extreme tilt, in which case the
corresponding dynamics is very special. But, from Table 1, there
do exist some extreme tilt sinks for $\gamma \leq \frac{4}{3}$.
For example, for Bianchi type VIII models there is a unique future
attractor ${\tilde{E}_1}$ which is of extreme tilt for all
$\gamma>1$, but from equation (\ref{28}) we see that there are
models with $ \frac{10}{9} < \gamma < \frac{4}{3}$ that lead to
$0<\geff<\frac{2}{3}$, and thus generically mimic the dynamics of
quintessence cosmological models. In the case of Bianchi models of
type V, for $ \frac{6}{5} < \gamma < \frac{4}{3}$ we again obtain
quintessential behaviour (a similar behaviour will occur in type
VI$_0$  and VI$_h$ models; however, in all of these cases the
extreme tilt sink is not the only future attractor). For certain
values of $\Sigma_+$ (see Table 1), the Bianchi type VII$_h$
models also lead to quintessence. We note that spatially
homogeneous and inhomogeneous models with a cosmological constant
cannot generically give rise to quintessential behaviour since
there appear to be no future attractors with extreme tilt when
$\gamma < \frac{4}{3}$ (although, of course, the models are
already inflating).

\section{Fluid frame variables}

Let us next determine the asymptotic behaviour of the tilting
models in the fluid frame variables using the boost formulae given
in Appendix B. In particular, we shall present the de Sitter limit
and the Bianchi type VIII future asymptotic limits as examples.

For the de Sitter limit, the asymptotic form is given in \cite{leuw}
(eqs. (3.22)-(3.28) and (3.43)). Note
that the asymptotic form for $v_\alpha$ in \cite{leuw}, eqs. (3.27) and
(3.28) can be simplified to
\be
    v^\alpha = v_0^\alpha + \bigO(e^{-r\tau})\ ,\quad
    1-v^2 = e^{-rt} [ (1-v^2)_0 + \bigO(e^{-r\tau}) ]\ ,
\ee where $r = 2(3\gamma-4)/(2-\gamma)$. That is, the
$\bigO(e^{-\tau})$ term is zero. The outline of the proof is as
follows; we choose a spatial frame such that $v^\alpha \rightarrow
(1,0,0)$ and we show that the $\bigO(e^{-\tau})$ term is zero. The
result is true for any Fermi-propagated frame.

With this simplification, we apply the boost formulae in the
appendix to the asymptotic forms to obtain
the fluid frame variables for the case $\gamma >
\frac43$:
\begin{align}
    \Hf &= \frac{2}{3(2-\gamma)}
        \Gamma_0 \sqrt{\frac{\Lambda}{3}} e^{\frac12r\tau}
        +\bigO( e^{-\frac12r\tau} + e^{(\frac12r-1)\tau} )
\\
    a_{\rm fluid}^\alpha &= - \Gamma_0 \sqrt{\frac{\Lambda}{3}}
                v_0^\alpha e^{\frac12r\tau}
    + [ \text{coef} ] e^{(r-1)\tau}
    +\bigO( 1 + e^{(\frac12r-1)\tau} )
\\
    \dot{u}_{\rm fluid}^\alpha
    &= \frac{2(\gamma-1)}{2-\gamma} \Gamma_0
    \sqrt{\frac{\Lambda}{3}} v_0^\alpha e^{\frac12r\tau}
    + [ \text{coef} ] e^{(r-1)\tau}
    +\bigO( 1 + e^{(\frac12r-1)\tau} )
\\
    \sigma_{\rm fluid}^{\alpha\beta}
    &= \frac12 r \Gamma_0 \sqrt{\frac{\Lambda}{3}}
    v_0^{\la\alpha} v_0^{\beta\ra} e^{\frac12r\tau}
    + [ \text{coef} ] e^{(r-1)\tau}
    +\bigO( 1 + e^{(\frac12r-1)\tau} )\ ,
\end{align}
all of which diverge as $\tau \rightarrow \infty$, while
\be
        \omega_{\rm fluid}^{\alpha\beta},\
        \Omega_{\rm fluid}^{\alpha\beta},\
        n_{\rm fluid}^{\alpha\beta}
    = \bigO(1 + e^{(r-1)\tau})\ ,
\ee
which diverge for $\gamma > \frac{10}{7}$.

The factors that contribute to the divergence (in the
$e^{\frac12r\tau}$ term) are the $\Gamma H$ term (affecting $\Hf$,
$a_{\rm fluid}^\alpha$ and $\dot{u}_{\rm fluid}^\alpha$), and the
$\mathbf{e}_0$ terms (affecting $\Hf$, $\dot{u}_{\rm
fluid}^\alpha$ and $\sigma_{\rm fluid}^{\alpha\beta}$). We now
look at the general situation for the existence of a singularity.
We observe that, in order for a singularity to occur along the
fluid congruence, the factor $\Gamma H$ in (\ref{def_T}) must
diverge sufficiently rapidly. The factor $\Gamma H$ also appears
in the boost formulae for $H$, $a_\alpha$ and $\dot{u}_\alpha$, so
these variables for the fluid frame generally diverge, unless any
cancellation occurs. This indicates that a singularity is
generally characterized by the divergence of $H$, $a_\alpha$ and
$\dot{u}_\alpha$. This leads us to speculate that a singularity
along a congruence is primarily affected by the rate of expansion
and the acceleration of the congruence rather than its shear and
vorticity. Furthermore, the spatial curvature component $a_\alpha$
plays a more significant role than $n_{\alpha\beta}$ and the frame
rotation $\Omega_{\alpha\beta}$ plays a less significant role.

We next present the asymptotic forms for the dominant variables
for the Bianchi type VIII limit. From \cite{HLim}, the asymptotic
forms for the variables along the congruence normal to the SH
hypersurfaces are
\begin{gather}
    H \approx H_0 e^{-\frac32\tau}\ ,\quad
    \sigma_+ \approx \tfrac12 H_0 e^{-\frac32\tau} \ ,\quad
    n_+ \approx (n_+)_0 \tau^{-\frac34}\ ,
\\
    1-v^2 \approx (1-v^2)_0 \tau^{\frac{1}{2-\gamma}}
        e^{-\frac{6(\gamma-1)}{2-\gamma}\tau}\ ,\quad
    v_1 \approx 1 + \bigO(1-v^2)\ ,\quad
    v_2, v_3 \approx \bigO(\tau^\frac34 e^{-\frac32\tau})
\\
    \sigma_{12}, \sigma_{13} \approx \bigO(\tau^{-\frac14} e^{-3\tau})
    \ ,\quad
    \sigma_1 \equiv (\sigma_-^2 + \sigma_\times^2 + n_-^2 +
    n_\times^2) H^2
    =\bigO(\tau^{-1})\ ,
\end{gather}
as $\tau \rightarrow \infty$, valid for $1 < \gamma <2$.
Applying the boost formulae in the appendix, we obtain
\begin{gather}
    \Hf \approx \frac{1}{2-\gamma} \Gamma_0
        H_0 \tau^{ - \frac{1}{2(2-\gamma)} }
        e^{ \frac{3(3\gamma-4)}{2(2-\gamma)} \tau}
\\
    a^1_{\rm fluid} \approx (2-\gamma) \Hf \ ,\quad
    \dot{u}^1_{\rm fluid} \approx (2\gamma-1) \Hf \ ,\quad
    (\sigma_+)_{\rm fluid} \approx (3\gamma-4) \Hf \ ,
\\
    (n_+)_{\rm fluid} \approx \Gamma_0 (n_+)_0
    \tau^{- \frac{8-3\gamma}{4(2-\gamma)} }
        e^{\frac{3(\gamma-1)}{2-\gamma}\tau} \ ,
\end{gather}
as $\tau \rightarrow \infty$, while other fluid frame variables are less
dominant.

As in the de Sitter example, for $\frac43 < \gamma < 2$,
 the factor $\Gamma H$ contributes to the
divergence of $\Hf$, $a^1_{\rm fluid}$ and $\dot{u}^1_{\rm fluid}$, while
the $\mathbf{e}_0$ terms contribute to the divergence of $\Hf$,
$\dot{u}^1_{\rm fluid}$ and $(\sigma_+)_{\rm fluid}$.
In addition, the diverging factor $\Gamma n_+$ contributes to the
diverging $(n_+)_{\rm fluid}$, and $\Gamma \sigma_+$ contributes to
$(\sigma_+)_{\rm fluid}$.

From the asymptotic behaviour of the extremely tilting models in
the fluid frame variables we have found that the length scale and
the Hubble parameter (i.e., the expansion) diverge in finite
proper time, and the singularity is consequently a future sudden
singularity in the sense of Barrow \cite{BarTsa}. In addition, the
shear and the acceleration of the fluid congruence also diverge
(while the energy density tends to zero). That is, the fluid
congruence experiences a 'finite kinematic' singularity, and
shares many of the physical properties of a big rip singularity
\cite{phantom}.  A 'finite kinematic' singularity is the analogue
of a 'conformal' singularity \cite{KingEllis,CollinsEllis} or a
'finite density' singularity  \cite{CollinsEllis} (in which the
energy density tends to a finite value) that occurs in the Bianchi
type V models. In particular, at a 'finite kinematic' singularity
all of the Weyl invariants remain bounded (although components of
the Weyl tensor can diverge).

\section{LRS Bianchi type V extendible solution}

As noted earlier, a similar situation can  occur to the past. As a
final example, let us consider the asymptotic dynamical behaviour
of a class of  Bianchi type V solutions. Collins and Ellis
\cite{CollinsEllis} showed that there exists a set of typical LRS
Bianchi V solutions that, into the past, can be extended along the
fluid congruence, from a spatially homogeneous region of spacetime
into a stationary region (see their figures 3--7). Along the
normal congruence, however, an initial singularity is reached,
with the limits \be
    (\Sp,A,v) \rightarrow (\Sp _0, 1+ \Sp _0, 1), ~~ -1<\Sp _0<0,
\ee and $H \approx H_0 e^{-(1-2\Sp _0)\tau}$,
$\Gamma \approx \Gamma_0 e^{-(1-2\Sp _0)\tau}$. The proper time
elapsed along the fluid congruence up to the Cauchy horizon is \be
    \Delta T = \int_{-\infty}^{\tau_0} \frac{1}{\Gamma H} \d\tau
    \approx \int_{-\infty}^{\tau_0} \frac{1}{\Gamma_0 H_0}
    e^{2(1-2\Sp _0)\tau} \d\tau\ ,
\ee which is finite.

Here, we apply the boost formulae to the asymptotic forms
to check that the fluid frame
variables are bounded at the Cauchy horizon, consistent with the
extendible solution:%
\footnote{The asymptotic forms for $\Sp$ and $A$ need to be
obtained explicitly up to the order $e^{2(1-2\Sp _0)\tau}$, and
$\Gamma$ up to the order $e^{(1-2\Sp _0)\tau}$, as cancellations occur
at leading orders to give the eventual bounded limits.  Heuristically, we obtain
the asymptotic forms as follows: the ansatz $\Sp = \Sp _0 + \Sp _1 e^{k_1 \tau} + ...$,
$A = 1 +  \Sp _0 + A _1 e^{k_2 \tau} + ...$, is substituted into the $\Sp$- and $A$-
evolution equations and the constraint equations to give $k_1=k_2=2(1-2\Sp _0)$,
and $\Sp _1$ and $A _1$ explicitly in terms of $\Sp _0$ and $\Gamma_0$. This ansatz is then
substituted into the $\Gamma$-evolution equation to give $\Gamma = \Gamma_0 
e^{-(1-2\Sp _0)\tau} +  \Gamma_1 e^{(1-2\Sp _0)\tau}$, where $\Gamma_1$ is given
explicitly in terms of  $\Sp _0$ and $\Gamma_0$.}
\begin{align}
    \Hf &= -\frac{(2-\gamma)\Sp _0^2
        +2\Sp _0-2\gamma}{3\gamma(2-\gamma)} \frac{H_0}{\Gamma_0}
        + \bigO(e^{(1-2\Sp _0)\tau})
\\
    (a_1)_{\rm fluid} &= \tfrac12 (1+\Sp _0)(1+ \Sp _0
            -\tfrac2\gamma\Sp _0) \frac{H_0}{\Gamma_0}
                + \bigO(e^{(1-2\Sp _0)\tau})
\\
    (\dot{u}_1)_{\rm fluid} &= -\frac{(\gamma-1)[(2-\gamma)\Sp _0^2
                +2\Sp _0-2\gamma]}{\gamma(2-\gamma)} \frac{H_0}{\Gamma_0}
                + \bigO(e^{(1-2\Sp _0)\tau})
\\
    (\sigma_+)_{\rm fluid} &= \frac{(3\gamma-4)(2-\gamma)\Sp _0^2
        -(6\gamma^2-18\gamma+8)\Sp _0 +
        \gamma(3\gamma-2)}{6\gamma(2-\gamma)} \frac{H_0}{\Gamma_0}
                + \bigO(e^{(1-2\Sp _0)\tau})\ ,
\end{align}
as $\tau \rightarrow - \infty$, while other components of the
fluid frame variables are identically zero.

Therefore, in this case, although the proper time as measured by
an observer is finite, there is a congruence along which the
evolution can be continued into the past. This raises the question
of whether, for any particular exact cosmological model or
corresponding asymptotic state, there exists a congruence whose
observers cross a Cauchy horizon into a space-time region not
accessible by the observers along the standard congruence of the
model considered. In particular,  it is of interest to determine
whether there is a congruence which can be continued beyond the
asymptotic state in any of the examples discussed above in which
an extreme tilt future asymptotic state is approached in finite
proper time.

\section{Discussion}

We have found that for $\gamma > \frac{4}{3}$ a 'finite kinematic'
singularity develops in which the nature of geometrically defined
surfaces change from spacelike to null in a finite time as
measured by the fluid. The field equations are structurally
unstable, and the question consequently arises as of whether these
models are mathematically well-defined. In any case, to fully
understand the behaviour of these models and their physical
properties, the dynamics need to be studied using a formulation
adapted to the fluid (i.e., utilizing a fluid-comoving frame).

On the other hand, this mathematical behaviour might lead to some
interesting physics. Expanding universes that come to a violent
end after a finite proper time have arisen in a different context.
Models with a constant equation of state parameter $\gamma<0$,
dubbed 'phantom energy', lead to a singularity called the big rip
\cite{phantom}. In this paradigm, during the cosmic evolution the
length scale grows more rapidly than the Hubble distance and
consequently blows up in a finite proper time, and is typically
characterized by a divergent pressure and acceleration. As the big
rip singularity is approached, both the strong and weak energy
conditions are violated. {\footnote{We are only considering
perfect fluid models here; it may be of interest to investigate
more realistic fluids with viscosity.}} The details of the
pathological behaviours depend on the particular phantom
cosmological model under consideration, and less violent
singularities such as future sudden singularities are possible
\cite{BarTsa}.

To establish the connection with phantom cosmology we determined
the effective equation of state in the models under consideration.
We found that as the asymptotic state is approached the
ultra-radiative perfect fluid effectively behaves like a phantom
energy in an isotropic and spatially flat spacetime. It is
important to note that the energy conditions of the perfect fluid
are nowhere violated. In order to compare the future asymptotic
behaviour with the behaviour of various types of big rip
singularities \cite{phantom} and future sudden singularities
\cite{BarTsa} we computed the future asymptotic dynamical
behaviour of the tilting models.

Let us discuss the physical consequences of this dynamical
behaviour in a little more detail. The models presumably spend a
period of time approximately isotropic (i.e., close to a flat
Friedmann saddle point), with a very small (but non-vanishing)
tilt. Thereafter, the models begin to evolve away from isotropy.
Since the tilt is non-zero, for $\gamma > \frac{4}{3}$ the models
generically evolve towards an asymptotic state with extreme tilt.
As the tilt becomes extreme, observers moving with the tilting
fluid will experience dynamical behaviour mimicking that of a
phantom cosmology. We note that such behaviour seems to be
consistent with galaxy, CMB and supernovae observational data
\cite{obs}. Moreover, unlike in conventional phantom cosmology, in
the models studied here there is no need for any exotic forms of
matter; conventional matter which is tilting suffices.%
\footnote{In a braneworld approach, accelerating universes can also result
without a cosmological constant or other form of dark energy
\cite{DGP}.}
Indeed, other pathologies, such as the existence of
ghosts, are avoided in the models described here.%
\footnote{This
is also the case in alternative models to phantom cosmology which
result from alternative theories of gravity, theories with
non-minimal couplings, and models in which the dark energy and
quintessence field interact \cite{inter}.}
In addition, as noted
above, due to the existence of future attractors with extreme tilt
the dynamical behaviour described here is generic.

We have also found that there are tilting SH perfect fluid models
with $\gamma < \frac{4}{3}$ which have extreme tilt asymptotically
and have an effective quintessence equation of state
($0<\geff<\frac{2}{3}$) \cite{quint}. Unlike the case of the
cosmologies with $\gamma > \frac{4}{3}$ discussed above in which
the phantom behaviour is generic, the quintessential behaviour is
typical (i.e., there are other future asymptotic behaviours
possible) and model dependent (i.e., it does not occur in all
Bianchi models). The details of the extreme tilt sinks in
different Bianchi models for $\gamma \leq \frac{4}{3}$ is given in
Table 1, and the corresponding ranges of $\gamma$ that lead to
$0<\geff<\frac{2}{3}$ and thus mimic the dynamics of quintessence
cosmological models is discussed in Section 5.

In future work we shall further study the physics of the models
with tilt, from the perspective of the observers moving with the
fluid matter congruence, using the formalism developed in this
paper. We note that the tilting models studied here agree with
observations in the same sense that the phantom cosmologies and
quintessence models agree with observations, and thus deserve
further scrutiny. In particular, we will investigate the effect of
tilt on cosmic microwave background radiation observations. It is
possible that these tilting models may offer an explanation for
the low quadrupole amplitude and other anomalies on large angular
scales \cite{Jaffe} found in the WMAP data \cite{Oliveira-Costa}.

\section*{Acknowledgments}
We thank Henk van Elst for helpful discussions on the boost
formulae. This work was supported by a Killam Postdoctoral
Fellowship (SH) and the Natural Sciences and Engineering Research
Council of Canada (AC).

\appendix
\section{LRS Bianchi type V models in fluid-frame variables}
\label{AppV} In the LRS Bianchi type V case, the evolution
equations in the fluid frame can be presented as a system of two
ODEs, simplified
from the equations for general spatially inhomogeneous models%
\footnote{See \cite{Henkweb} for the general equations in
component form.} by setting
\begin{gather}
        \parb_1 = v \parb_0\, ,\quad
        \parb_2=\parb_3=0\, ,\quad
        \Sigma_{\alpha\beta} = \text{diag}(-2\Sigma_+,\Sigma_+,\Sigma_+)
        \, ,\quad
\\
        N_{\alpha\beta}=0\, ,\quad
        A_\alpha = (A,0,0)\, ,\quad
        r_\alpha = (r,0,0)\, ,\quad
        \dot{U}_\alpha = (\dot{U},0,0)\, ,\quad
\\
        R_\alpha=W_\alpha=0\, ,\quad
        P = 0\, ,\quad Q_\alpha=0\, ,\quad \Pi_{\alpha\beta}=0\, .
\end{gather}
The resulting system is
\begin{align}
        v' &= (2\Sigma_+ + 3\gamma-4) v
\\
        \Sigma_+' &= - 3(1+\Sigma_+) G_-^{-1} [ G_+ \Sigma_+ +
                        2(\gamma-1)^2 v^2]
\notag\\
        &\qquad -\frac{A}{3vG_-} [ (9+v^2)\Sigma_+^2 +
                2(6\gamma-5)v^2\Sigma_+ + (3\gamma-2)v^2 ]\, ,
\end{align}
where
\begin{align}
        A &= \frac{-b - \sqrt{b^2-4ac}}{2a}
\\
        a &= -\tfrac12(3\gamma-2)
\\
        b &= [-3G_- + (2+G_-)(1+\Sigma_+)] /v
\\
        c &= \tfrac32(2-\gamma)(1+\Sigma_+)^2
                + 3(\gamma-1)G_-(1+\Sigma_+)
\\
        G_\pm &= 1 \pm (\gamma-1)v^2\, .
\end{align}
The other fluid-frame variables are given by
\begin{align}
        \dot{U} &= 3(\gamma-1)v
\\
        \frac{r}{v} &= q+1 = -\frac{3}{G_-} [ G_+ \Sigma_+ + 2(\gamma-1)^2
                v^2 ]
                - \frac{A}{3vG_-}[ (9+v^2)\Sigma_+ + (3\gamma-2)v^2 ]
\\
        K &= 2(\gamma-1)(1+\Sigma_+)v^2+A^2+\tfrac23v(1+\Sigma_+)A
\\
        \Omega &= 1- \Sigma_+^2 - K\, .
\end{align}

On approach to the $M^-$ point, $v$ tends to $-1$ and the
fluid-frame variables are: \be
        \Sigma_+ = -\tfrac12(3\gamma-4) \, ,\quad
        A = \tfrac32(2-\gamma) \, ,\quad
        \dot{U} = -3(\gamma-1) \, ,\quad
        r = \tfrac32(3\gamma-4)\, ,
\ee which lead to \be
        \Omega=P=0\, ,\quad
        \Sigma^2 = \tfrac14(3\gamma-4)^2\, ,\quad
        K = \tfrac34(3\gamma-2)(2-\gamma)\, ,\quad
        q = -\tfrac12(9\gamma-10)\, .
\ee

Many of the calculations are more straightforward in fluid frame
variables. For example, using these equations we can easily
calculate the asymptotic decay rates (10) and (11) on the approach
to $M^-$. We can also calculate the effective equation of state
(assuming that the effective solution is spatially flat and
isotropic, so that $K_{\rm eff} = 0 = \Sigma_{\rm eff}$, 
$\Omega_{\rm eff} = 1$ and $q_{\rm fluid} = \tfrac12 (\Omega+3P)_{\rm eff}$):
\be
       \gamma_{\rm eff} = - (3\gamma-4) \,\ee
which is negative for $\gamma > \tfrac43$ on approach to $M^-$.

\section{Boost Formulae}
{\large {\allowdisplaybreaks The commutation functions are given
by:\be
    [ \mathbf{e}_a , \mathbf{e}_b ] = \gamma^c{}_{ab} \mathbf{e}_c
    \,
\ee which can be decomposed in the following way:
\begin{align}
    \gamma^0{}_{0\a} &= \udot_\a
\\
    \gamma^\b{}_{0\a} &= - H \delta_\a{}^\b - \sigma_\a{}^\b
                - \omega_\a{}^\b - \Omega_\a{}^\b
\\
    \gamma^0{}_\ab &= 2 \omega_\ab
\\
    \gamma^\mu{}_\ab &= \eps_{\ab\nu} n^{\mu\nu}
        + a_\a \delta_\b{}^\mu - a_\b \delta_\a{}^\mu
\end{align}
where $\sigma_\ab$ is symmetric and traceless; $n_\ab$ is
symmetric; $\omega_\ab$ and $\Omega_\ab$ are antisymmetric. For
convenience, we also give the expressions for the components in
terms of the commutation functions (angled brackets denote
trace-free symmetrization):
\begin{gather}
    H = -\tfrac13 \gamma^\mu{}_{0\mu}
    \ ,\quad
    \sigma_\ab = \gamma_{\la \b 0 \a \ra}
    \ ,\quad
    \omega_\ab + \Omega_\ab = \gamma_{[ \b 0 \a ]}
\\
    n^\ab = \tfrac12 \eps^{\mu\nu(\b} \gamma^{\a)}{}_{\mu\nu}
    \ ,\quad
    a_\a = -\tfrac12 \gamma^\mu{}_{\mu\a}\ .
\end{gather}
The boost formulae for the orthonormal frame vector fields are:
\begin{align}
    \ehat_0 &= \G \mathbf{e}_0 + \Gv^\mu \mathbf{e}_\mu
\\
    \ehat_\a &= \Gv_\a \mathbf{e}_0
        + B_\a{}^\mu \mathbf{e}_\mu
\end{align}
where \be
    B_\a{}^\mu = \left[ \delta_\a{}^\mu
                + \frac{\G^2}{\G+1} v_\a v^\mu \right]
    \ ,\quad
    \G = \frac{1}{\sqrt{1-v^2}}
    \ ,\quad
    v^2 = v_\mu v^\mu
    \ .
\ee The corresponding boost formulae for the commutation functions
are:
\begin{align}
    \hat{\gamma}^0{}_{0\a}
    &= \G [ \ehat_0(\Gv_\a)-\ehat_\a(\G) ]
    - \Gv_\mu [ \ehat_0(B_\a{}^\mu) - \ehat_\a (\Gv^\mu) ]
\notag\\
    &\quad
    + (\G B_\a{}^\mu - \G^2 v_\a v^\mu)
    ( \G \gamma^0{}_{0\mu} - \Gv_\nu \gamma^\nu{}_{0\mu} )
\notag\\
        &\quad
    + \G v^\mu B_\a{}^\nu ( \G \gamma^0{}_{\mu\nu}
        - \Gv_\z \gamma^\z{}_{\mu\nu} )
\\
    \hat{\gamma}^\b{}_{0\a}
    &= -\Gv^\b [ \ehat_0(\Gv_\a)-\ehat_\a(\G) ]
    + B_\mu{}^\b [ \ehat_0(B_\a{}^\mu) - \ehat_\a (\Gv^\mu) ]
\notag\\
        &\quad
    + (\G B_\a{}^\mu - \G^2 v_\a v^\mu)
    ( - \Gv^\b \gamma^0{}_{0\mu} + B_\nu{}^\b \gamma^\nu{}_{0\mu} )
\notag\\
        &\quad
    + \Gv^\mu B_\a{}^\nu ( - \Gv^\b \gamma^0{}_{\mu\nu}
        +  B_\z{}^\b \gamma^\z{}_{\mu\nu} )
\\
    \hat{\gamma}^0{}_\ab
    &= 2 \G \ehat_{[\a} (\Gv_{\b]})
        - 2 \Gv_\mu \ehat_{[\a}( B_{\b]}{}^\mu )
\notag\\
        &\quad
    + 2 \Gv_{[\a} B_{\b]}{}^\mu ( \G \gamma^0{}_{0\mu}
        - \Gv_\nu \gamma^\nu{}_{0\mu} )
\notag\\
        &\quad
    + B_\a{}^\mu B_\b{}^\nu ( \G \gamma^0{}_{\mu\nu}
        - \Gv_\z \gamma^\z{}_{\mu\nu} )
\\
    \hat{\gamma}^\gamma{}_\ab
    &= - 2 \Gv^\gamma \ehat_{[\a} (\Gv_{\b]})
        + 2 B_\mu{}^\gamma \ehat_{[\a}( B_{\b]}{}^\mu )
\notag\\
        &\quad
    + 2 \Gv_{[\a} B_{\b]}{}^\mu ( - \Gv^\gamma \gamma^0{}_{0\mu}
        + B_\nu{}^\gamma \gamma^\nu{}_{0\mu} )
\notag\\
        &\quad
        + B_\a{}^\mu B_\b{}^\nu ( - \Gv^\gamma \gamma^0{}_{\mu\nu}
                + B_\z{}^\gamma \gamma^\z{}_{\mu\nu} )
\end{align}
The corresponding boost formulae for the components of the
commutation functions are then:
\begin{align}
    \hat{H} &=
    \frac13 \left[ \ehat_\mu(\Gv^\mu)
            - \frac{\G}{\G+1} v^\mu \ehat_\mu(\G)
            +3\G H  - 2 \G a_\mu v^\mu + \G \udot_\mu v^\mu
    \right]
\\
    \hat{\sigma}_\ab &=
    \ehat_{\la\a}(\Gv_{\b\ra})
    - \frac{\G}{\G+1}v_{\la\b}\ehat_{\a\ra}(\G)
\notag\\
    &\quad
    + \G \sigma_\ab
    + \frac{\G^4}{(\G+1)^2} \left[v^2 v^\mu \sigma_{\mu\la\a}v_{\b\ra}
    - \sigma_{\mu\nu} v^\mu v^\nu v_{\la\a} v_{\b\ra} \right]
\notag\\
    &\quad
    - \G^2 v^\mu (\omega_{\mu\la\a} + \Omega_{\mu\la\a}) v_{\b\ra}
    + 2 \G^2 v^\mu \omega_{\mu\la\a} v_{\b\ra}
\notag\\
        &\quad
    + \G v^\mu \eps_{\mu\nu\la\a} n_{\b\ra}{}^\nu
    + \frac{\G^3}{\G+1} v^\mu \eps_{\mu\nu\la\a}v_{\b\ra}n^{\nu\z}v_\z
\notag\\
        &\quad
    + \G^2 a_{\la\a} v_{\b\ra}
    -\frac{\G^3}{\G+1} a_\mu v^\mu v_{\la\a} v_{\b\ra}
\notag\\
        &\quad
        + \G^2 \udot_{\la\a} v_{\b\ra}
        -\frac{\G^3}{\G+1} \udot_\mu v^\mu v_{\la\a} v_{\b\ra}
\\
    \hat{\omega}_\ab + \hat{\Omega}_\ab
    &= \ehat_{[\a}(\Gv_{\b]})
        - \frac{\G}{\G+1}v_{[\b}\ehat_{\a]}(\G)
    + \frac{2\G}{\G+1} v_{[\b} \ehat_0(\Gv_{\a]})
\notag\\
        &\quad
    + \G^2 v^\mu \sigma_{\mu[\a} v_{\b]}
\notag\\
        &\quad
    + \G(\omega_\ab +\Omega_\ab)
    + \frac{\G^4 v^2}{(\G+1)^2} v^\mu
        (\omega_{\mu[\a}+\Omega_{\mu[\a}) v_{\b]}
    + 2 \G^2 v^\mu \omega_{\mu[\a} v_{\b]}
\notag\\
        &\quad
        + \G v^\mu \eps_{\mu\nu[\a} n_{\b]}{}^\nu
        + \frac{\G^3}{\G+1} v^\mu \eps_{\mu\nu[\a}v_{\b]}n^{\nu\z}v_\z
\notag\\
        &\quad
        + \G^2 a_{[\a} v_{\b]}
    + \G^2 \udot_{[\a} v_{\b]}
\\
    \hat{\omega}_\ab
    &= \ehat_{[\a}(\Gv_{\b]})
        - \frac{\G}{\G+1}v_{[\b}\ehat_{\a]}(\G)
\notag\\
        &\quad
    - \G^2 v^\mu \sigma_{\mu[\a} v_{\b]}
\notag\\
        &\quad
    + \G^2 v^\mu (\omega_{\mu[\a}+\Omega_{\mu[\a}) v_{\b]}
    + \G \omega_\ab - \frac{2\G^3}{\G+1}  v^\mu \omega_{\mu[\a} v_{\b]}
\notag\\
    &\quad
    -\tfrac12 \G \eps_{\a\b\mu} n^{\mu\nu} v_\nu
        - \frac{\G^3}{\G+1} v^\mu \eps_{\mu\nu[\a}v_{\b]}n^{\nu\z}v_\z
\notag\\
        &\quad
        - \G^2 a_{[\a} v_{\b]}
        - \G^2 \udot_{[\a} v_{\b]}
\\
    \hat{n}_\ab
    &=
    \frac{\G}{\G+1} \eps^{\mu\nu}{}_{(\a|} v_\nu \ehat_\mu(\Gv_{|\b)})
    - \frac{\G}{\G+1} \eps^{\mu\nu}{}_{(\a} v_{\b)} \ehat_\mu(\Gv_\nu)
\notag\\
        &\quad
    - \G v^\mu \eps_{\mu\nu(\a} \sigma_{\b)}{}^\nu
    - \frac{\G^3}{\G+1} v^\mu \eps_{\mu\nu(\a} v_{\b)}\sigma^{\nu\z}v_\z
\notag\\
        &\quad
    - \G v^\mu \eps_{\mu\nu(\a}(\omega_{\b)}{}^\nu+\Omega_{\b)}{}^\nu)
        - \frac{\G^3}{\G+1} v^\mu \eps_{\mu\nu(\a} v_{\b)}
        (\omega^{\nu\z}+\Omega^{\nu\z}) v_\z
\notag\\
        &\quad
    - \G \eps_{\mu\nu(\a} v_{\b)} \omega^{\mu\nu}
    + \frac{2\G^3}{\G+1} v^\mu \eps_{\mu\nu(\a} v_{\b)} \omega^{\nu\z} v_\z
\notag\\
        &\quad
    + \G n_\ab + \frac{\G^4}{(\G+1)^2} \left[ v^2 v^\mu n_{\mu(\a}
    v_{\b)} - n_{\mu\nu} v^\mu v^\nu v_\a v_\b \right]
\notag\\
        &\quad
    -\G^2 v^\mu \eps_{\mu\nu(\a} v_{\b)} a^\nu
    -\G^2 v^\mu \eps_{\mu\nu(\a} v_{\b)} \udot^\nu
\\
    2\hat{a}_\a &=
    \frac{\G}{\G+1} v^\mu \ehat_\mu(\Gv_\a)
    - \frac{\G}{\G+1} v_\a \ehat_\mu(\Gv^\mu)
\notag\\
        &\quad
    -2\G H v_\a
    + \G^2 v^\mu \sigma_{\a\mu}
    - \frac{\G^3}{\G+1} \sigma_{\mu\nu} v^\mu v^\nu v_\a
\notag\\
    &\quad
    + \G^2 v^\mu (\omega_{\a\mu}+\Omega_{\a\mu})
    - 2\G^2 v^\mu \omega_{\a\mu}
\notag\\
        &\quad
    + \G^2 v^\mu \eps_{\mu\nu\a} n^{\nu\z} v_\z
\notag\\
        &\quad
    + (2+\G^2v^2) a_\a - \frac{\G^4v^2}{(\G+1)^2} a_\mu v^\mu v_\a
\notag\\
        &\quad
    + \G^2 v^2 \udot_\a - \G^2 \udot_\mu v^\mu v_\a
\\
    \hat{\udot}_\a
    &= \ehat_0(\Gv_\a) - \frac{\G}{\G+1} v_\a \ehat_0(\G)
\notag\\
        &\quad
    + \G H v_\a
    + \G^2 v^\mu \sigma_{\a\mu}
        - \frac{\G^3}{\G+1} \sigma_{\mu\nu} v^\mu v^\nu v_\a
\notag\\
        &\quad
        + \G^2 v^\mu (\omega_{\a\mu}+\Omega_{\a\mu})
        - 2\G^2 v^\mu \omega_{\a\mu}
\notag\\
        &\quad
        + \G^2 v^\mu \eps_{\mu\nu\a} n^{\nu\z} v_\z
        + \G^2 v^2 a_\a - \G^2 a_\mu v^\mu v_\a
\notag\\
        &\quad
    +\G^2 \udot_\a - \frac{\G^3}{\G+1} \udot_\mu v^\mu v_\a
\end{align}

The energy-momentum tensor $T_{ab}$ has a standard decomposition
relative to a congruence of world-lines: \be
    T_{00} = \mu\ ,\quad
    T_{0\a} = - q_\a \ ,\quad
    T_{\a\b} = p \delta_{\a\b} + \pi_{\a\b}\ ,
\ee where $\mu$ is the total energy density, $q_\a$ is the energy
flux, $p$ is the isotropic pressure and $\pi_{\a\b}$ is the
trace-free anisotropic pressure, all relative to the above
congruence. These variables are called the source terms.

Boosting the orthonormal frame results in the following boost
formulae for the source terms:
\begin{align}
    \hat{\mu} &= \mu + \G^2 v^2 (\mu+p)
        - 2\G^2 q_\mu v^\mu + \G^2 \pi_{\mu\nu} v^\mu v^\nu
\\
    \hat{p} &= p + \tfrac13 \G^2 v^2 (\mu+p)
        - \tfrac23 \G^2 q_\mu v^\mu
        + \tfrac13 \G^2 \pi_{\mu\nu} v^\mu v^\nu
\\
    \hat{q}_\a &= - \G^2 (\mu+p) v_\a
        + \G q_\a + \frac{\G^2(2\G+1)}{\G+1} q_\mu v^\mu v_\a
\notag\\
    &\quad
        - \G v^\mu \pi_{\mu\a}
        - \frac{\G^3}{\G+1} \pi_{\mu\nu} v^\mu v^\nu v_\a
\\
    \hat{\pi}_{\a\b} &= \G^2 (\mu+p) v_{\la\a} v_{\b\ra}
        - 2 \G v_{\la\a} q_{\b\ra}
        - \frac{2\G^3}{\G+1} q_\mu v^\mu v_{\la\a} v_{\b\ra}
\notag\\
    &\quad
        + \pi_{\a\b}
        + \frac{2\G^2}{\G+1} v^\mu \pi_{\mu\la\a} v_{\b\ra}
+ \frac{\G^4}{(\G+1)^2} \pi_{\mu\nu} v^\mu v^\nu v_{\la\a}
v_{\b\ra}
\end{align}

} }

\bibliographystyle{amsplain}

\end{document}